\documentclass[english,aps,floatfix]{revtex4}
\usepackage{graphicx}
\usepackage{dsfont}
\usepackage{enumerate}
\usepackage{slashed}
\usepackage{amsfonts}
\usepackage[dvips]{epsfig}
\usepackage{babel}
\usepackage[dvips]{color}
 \usepackage{amsmath, amsthm, amssymb}
\begin{document}

\title{Transition Decomposition of Quantum Mechanical Evolution}

\author{Y. Strauss\textsuperscript{1}, J. Silman\textsuperscript{2,3}, S. Machnes\textsuperscript{2,4}, and L.P.
Horwitz\textsuperscript{2,5,6}}

\affiliation{\textsuperscript{1}Einstein Institute of Mathematics, Edmond J. Safra campus,
The Hebrew University of Jerusalem, Jerusalem 91904, Israel\\
\textsuperscript{2}School of Physics and Astronomy, Raymond and Beverly Sackler
Faculty of Exact Sciences, Tel-Aviv University, Tel-Aviv 69978, Israel\\
\textsuperscript{3}Laboratoire d'Information Quantique, Universit\'{e} Libre de
Bruxelles, 1050 Bruxelles, Belgium\\
\textsuperscript{4}Institut f\"{u}r Theoretische Physik, Universit\"{a}t
Ulm, 89069 Ulm, Germany\\
\textsuperscript{5}Physics Department of Physics, Bar-Ilan University, Ramat-Gan
52900, Israel\\
\textsuperscript{6}Department of Physics, The Ariel University Center of Samaria,
Ariel 40700, Israel}

\begin{abstract}\textbf{
We show that the existence of the family of self-adjoint Lyapunov operators introduced in [J. Math. Phys. \textbf{51}, 022104 (2010)] allows for the decomposition of the state of  a quantum mechanical system into two parts: A past time asymptote, which is asymptotic to the state of the system at $t\rightarrow -\infty$ and vanishes at $t\rightarrow\infty$, and a future time asymptote, which is asymptotic to the state of the system at $t\rightarrow\infty$ and vanishes at $t\rightarrow-\infty$. We demonstrate the usefulness of this decomposition for the description of resonance phenomena by considering the resonance scattering of a particle off a square barrier potential. We show that the past time asymptote captures the behavior of the resonance. In particular, it exhibits the expected exponential decay law and spatial probability distribution.}
\end{abstract}
\maketitle

\section{Introduction}

In standard non-relativistic quantum mechanics time enters as a parameter,
external to the quantum system being investigated, and as such is
not a dynamical variable associated in any way with the system's dynamics.
However, there are many cases, such as experiments measuring the time
of arrival of particles at a detector, the decay time of an unstable
quantum system etc., which call for a more dynamical point of view
with respect to time. One approach to this problem consists of
the construction of time operators through the use of covariant positive
operator valued measures (POVMs) \cite{Holevo, Busch}.
Since the time operators built via the use of POVMs are in general
maximally symmetric, non-self-adjoint operators, the construction of
these objects bypasses an old theorem of Pauli \cite{Pauli}, stating that there
does not exist a self-adjoint time operator $T$ canonically conjugate
to a Hamiltonian $H$ whose spectrum is semibounded, such that the
pair $T$ and $H$ forms together an imprimitivity system \cite{Mackey}. However, in comparison to standard self-adjoint quantum
observables these operators present some difficulties, e.g., in the
relation between the algebra they generate and their spectral representations.

The dynamical role of time within the framework of standard quantum mechanics has recently been considered
in \cite{Yossi, Comptes Rendus} through the construction of a family of self-adjoint
Lyapunov operators.
Throughout this text we define a Lyapunov operator as a self-adjoint
operator whose expectation value is monotonically decreasing in time.
More precisely, let $\mathcal{H}$ be a Hilbert space corresponding
to a given quantum mechanical system, let $H$ be a self-adjoint Hamiltonian
generating its evolution, and let $\left|\psi(t)\right\rangle =U(t)\left|\psi\right\rangle =\exp(-iHt)\left|\psi\right\rangle $
be the state of the system at time $t$. Define the trajectory $\Psi_{\psi}$,
corresponding to an initial state $\left|\psi\right\rangle \in\mathcal{H}$,
as the set of states \begin{equation}
\Psi_{\psi}:=\left\{ \left|\psi(t)\right\rangle \right\} _{t\in\mathbb{R}^{+}}=\left\{ U(t)\left|\psi\right\rangle \right\} _{t\in\mathbb{R}^{+}}\,.\label{eq:recurring_traj}\end{equation}
Then the definition of a Lyapunov operator is as follows \cite{Yossi}:\\
\\
\textbf{Definition 1:} \emph{Let $M$ be a bounded self-adjoint operator on
$\mathcal{H}$. Let $\Psi_{\psi}$ be a trajectory corresponding
to an arbitrarily chosen initial state $\vert \psi \rangle \in\mathcal{H}$.
Denote by $M\left(\Psi_{\psi}\right)=\left\{ \langle \varphi\vert M \vert \varphi\rangle \mid \vert \varphi \rangle \in\Psi_{\psi}\right\} $
the collection of all expectation values of $M$ for normalized states
in $\Psi_{\psi}$. Then $M$ is a forward Lyapunov operator if
the mapping $\tau_{M,\,\psi}:\mathbb{R}^{+}\mapsto M(\Psi_{\psi})$
defined by \begin{equation}
\tau_{M,\,\psi}\left(t\right)=\left\langle \psi(t)\left|M\right|\psi(t)\right\rangle \end{equation}
is monotonically decreasing in time.}\hfill{}$\square$\\
\\
\textbf{Remark 1:} If in the definition above we require that $\tau_{M,\psi}$
be monotonically increasing instead of monotonically decreasing we
also obtain a valid definition of a Lyapunov operator. The requirement
that $\tau_{M,\psi}$ is monotonically decreasing is made purely
for the sake of convenience.
\bigskip{}

It follows from this definition that a Lyapunov operator allows for
the temporal ordering of states in any trajectory $\Psi_{\psi}$
 according to the ordering of the expectation values
in $M(\Psi_{\psi})$, thereby introducing temporal ordering into the Hilbert space $\mathcal{H}$
of any problem for which such an operator can be constructed.

The construction
of Lyapunov operators constitutes a somewhat conservative approach
to the problem of time in quantum mechanics. In general, Lyapunov
operators indicate the direction of flow of time in a system but
do not serve as time operators and are not intended for answering
questions of direct time measurements. Rather, the construction of
Lyapunov operators, and in particular the family of Lyapunov operators introduced
in \cite{Yossi, Comptes Rendus}, was carried out with a different
goal in mind. The temporal ordering introduced into the Hilbert space
is a fundamental property of a quantum mechanical problem admitting
 Lyapunov operators. A question then arises as to the possible implications this property may have on the
description of the dynamics of the system, i.e. is
there a way to `inject' the direction of time ordering introduced by the Lyapunov
operator into the description of the dynamics of a quantum system
in such a way that certain processes in the evolution of the system
obtain simple descriptions amenable to thorough analyses? The study
of such consequences of the existence of a Lyapunov operator has been
addressed to some extent in \cite{Yossi, Comptes Rendus}.
In the present paper we continue this line of investigation, and demonstrate the beginning of an affirmative answer
to the question raised above. Specifically, we show that the existence
of the family of Lyapunov operators introduced in \cite{Yossi, Comptes Rendus} leads to a certain description of the scattering process,
which we shall term \emph{the transition decomposition}, that is particularly
useful for treating the evolution of scattering resonances.
\section{Lyapunov operators}
In this section we state some key results from \cite{Yossi, Comptes Rendus}, which will provide the basis for the all that follows. The following theorem is an adaptation of results stated in \cite{Yossi, Comptes Rendus}:\\
\\
\textbf{Theorem 1:} \label{thm:forward_lypunov_op}\emph{Let $\mathcal{H}$
be the Hilbert space representing some quantum mechanical system,
let $H$ be its Hamiltonian, and let the spectrum of $H$
be $\mathbb{R}^{+}$, absolutely continuous, and uniformly degenerate.
Then for any choice of generalized eigenbasis $\left|E,\,\lambda\right\rangle $
of $H$, $\lambda$ being the degeneracy index, the operator \begin{equation}
M=-\frac{1}{2\pi i}\sum_{\lambda}\int_{0}^{\infty}dE\int_{0}^{\infty}dE'\left|E,\,\lambda\right\rangle \frac{1}{E-E'+i0^{+}}\left\langle E',\,\lambda\right|\,,\label{M_F}\end{equation}
where the summation over $\lambda$ may also stand for integration,
is a Lyapunov operator in the sense of Definition 1, 
and, moreover, for any state $\vert \psi \rangle \in \mathcal{H}$ \begin{equation}
\lim_{t\to\infty}\left\langle \psi\left(t\right)\left|M\right|\psi\left(t\right)\right\rangle =0\,,\qquad\lim_{t\to-\infty}\left\langle \psi\left(t\right)\left|M\right|\psi\left(t\right)\right\rangle =1\,.
\end{equation}}

\hfill{}$\square$\\
\textbf{Remark 2:} In this paper we shall consider only Hamiltonians for which the spectrum is $\mathbb R^+$, 
absolutely continuous, and uniformly degenerate. The results below apply also to the more general case of Hamiltonians for which the absolutely continuous 
spectrum is $\mathbb R^+$  and is uniformly degenerate. In such cases the constructions below hold in the subspace $\mathcal{H}_{ac}$.\\ 
\\
\textbf{Remark 3:} Note that each complete set of generalized eigentates of $H$
has associated with it a different Lypaunov operator, and as such Theorem 1 represents a prescription for the construction of a family of self-adjoint Lyapunov operators. In particular,
this still holds true for two sets that differ only by an energy dependent
phase. When discussing general properties of the set of Lypaunov operators
defined above, we shall often refer to $M$ without specifying the
underlying complete set of generalized eigenstates.\\
\\
\textbf{Remark 4:} The above family of self-adjoint Lyapunov operators has recently been generalized in \cite{Muga}.\\
\\
\textbf{Remark 5:} Throughout this paper we employ natural units, i.e. $\hbar = c = 1$.\\

\par The following corollary to Theorem 1 is crucial for what follows.\\
\\
\textbf{Corollary 1:} \emph{Let $\Lambda:=M^{1/2}$. Then for any state $\vert \psi \rangle \in \mathcal{H}$
\begin{equation}
\lim_{t\to\infty}\left\langle \psi\left(t\right)\left|\Lambda\right|\psi\left(t\right)\right\rangle =0\,,\qquad\lim_{t\to-\infty}\left\langle \psi\left(t\right)\left|\Lambda\right|\psi\left(t\right)\right\rangle =1\,.
\end{equation}}
\hfill{}$\square$
\section{The transition decomposition} 
In this section we present a decomposition of the wave-function into two components -- one that is asymptotic to the state of the system in the far past and vanishes in the far future, and one that is asymptotic to the state of the system in the far future and vanishes in the far past -- and a corresponding decomposition for operators when working in the Heisenberg picture. The transition over time between the two asymptotic
components suggests the possibility that this decomposition yields an appropriate
description of transient phenomena in the evolution of a quantum system.
This expectation is realized when we apply the decomposition to scattering problems, where it will be seen to afford
a highly useful description of the evolution of scattering resonances. This
task is taken up in the next section.\\

Let $\left|\psi(t)\right\rangle =U(t)\left|\psi\right\rangle $,
$\left|\psi\right\rangle \in\mathcal{H}$, $t\in\mathbb{R}$. At any moment in time $t$ express $\psi (t)$ as the sum of two components as follows:  

\begin{equation}
\vert\psi(t)\rangle=\vert\psi_{b}(t)\rangle+\vert\psi_{f}(t)\rangle\,,\label{eq:transition representation_schrodinger}\end{equation}
 where \begin{equation}
\vert\psi_{b}(t)\rangle :=\Lambda\left|\psi(t)\right\rangle \,,\qquad\vert\psi_{f}(t)\rangle :=(I-\Lambda)\left|\psi(t)\right\rangle \,.\end{equation}
The limits below are then an immediate consequence of Theorem 1 and Corollary 1:\begin{eqnarray}
\lim_{t\to-\infty}\bigl\Vert\left|\psi(t)\right\rangle -\vert\psi_{b}(t)\rangle \bigr\Vert=0\,, & & \quad\lim_{t\to\infty}\bigl\Vert\vert\psi_{b}(t)\rangle \bigr\Vert=0\,,\label{eq:backward_component_limits}\\
\lim_{t\to-\infty}\bigl\Vert\vert\psi_{f}(t)\rangle \bigr\Vert=0\,, &  & \quad\lim_{t\to\infty}\bigl\Vert\vert\psi(t)\rangle -\vert\psi_{f}(t)\rangle \bigr\Vert=0\,.\label{eq:forward_component_limits}\end{eqnarray}
Eqs. (\ref{eq:transition representation_schrodinger}), (\ref{eq:backward_component_limits}), and
(\ref{eq:forward_component_limits}) imply that $\left|\psi(t)\right\rangle $
can be decomposed into a sum of two components, $\left|\psi_{b}(t)\right\rangle $
and $\vert\psi_{f}(t)\rangle $, such that $\vert\psi_{b}(t)\rangle $
vanishes in the future time asymptote and is asymptotic to $\left|\psi(t)\right\rangle $
in the past time asymptote, and $\vert\psi_{f}(t)\rangle $
vanishes in the past time asymptote and is asymptotic to $\left|\psi(t)\right\rangle $
in the future time asymptote. We refer to $\vert\psi_{b}(t)\rangle $
as the \emph{backward asymptotic component} and to $\vert\psi_{f}(t)\rangle $
as the \emph{forward asymptotic component} of $\left|\psi(t)\right\rangle $.
Since the decomposition here is of the evolving state $\left|\psi(t)\right\rangle $,
we call the decomposition in Eq. (\ref{eq:transition representation_schrodinger})
the \emph{transition decomposition in the Schr\"{o}dinger picture.} In
this decomposition the evolution of $\vert\psi(t)\rangle $
is represented as a transition from its backward asymptotic component
to its forward asymptotic component.\\

The transition decomposition in the Schr\"{o}dinger picture, Eq.
(\ref{eq:transition representation_schrodinger}), gives rise to a corresponding
decomposition in the Heisenberg picture. Let $X$
be a self-adjoint operator representing some physical observable and
let $X\left(t\right)=U^{\dagger}\left(t\right)XU\left(t\right)$ be
its Heisenberg evolution. Consider the expectation value of
$X(t)$ for some arbitrary state $\left|\psi\right\rangle \in\mathcal{H}$
and apply the decomposition in Eq. (\ref{eq:transition representation_schrodinger}).
We have \begin{eqnarray}
& \left\langle \psi\left|X\left(t\right)\right|\psi\right\rangle  =  \left\langle \psi\left(t\right)\left|X\right|\psi\left(t\right)\right\rangle  =  \langle \psi_{b}\left(t\right)+\psi_{f}\left(t\right)\left|X\right|\psi_{b}\left(t\right)+\psi_{f}\left(t\right)\rangle & \nonumber \\ \label{eq:schrodinger_to_heisenberg_trans_decomp}
 &   =\langle \psi_{b}\left(t\right)\left|X\right|\psi_{b}\left(t\right)\rangle +\bigl(\langle \psi_{b}\left(t\right)\left|X\right|\psi_{f}\left(t\right)\rangle +\langle \psi_{f}\left(t\right)\left|X\right|\psi_{b}\left(t\right)\rangle \bigr)+\langle \psi_{f}\left(t\right)\left|X\right|\psi_{f}\left(t\right)\rangle \,.&\end{eqnarray}
Define now the following decomposition of $X$, which will be seen below to correspond to the three different
terms on the right-hand side of Eq. (\ref{eq:schrodinger_to_heisenberg_trans_decomp}),
\begin{equation}
X:=X_{b}+X_{tr}+X_{f}\,,\label{eq:transition_representation_heisenberg}\end{equation}
with \begin{equation}
X_{b}  :=  \Lambda X\Lambda\,,\qquad X_{tr}:=\Lambda X(I-\Lambda)+(I-\Lambda)X\Lambda\,,\qquad X_{f} :=  (I-\Lambda)X(I-\Lambda)\,.\end{equation}
Indeed, if $X_{b}(t)$, $X_{tr}(t)$, and $X_{f}(t)$, respectively, are
the Heisenberg evolutions of $X_{b}$, $X_{tr}$, and $X_{f}$, so that
\begin{equation}X(t)=X_{b}(t)+X_{tr}(t)+X_{f}(t)\, \label{eq:trans_rep_heisenberg_t}\,, \end{equation} then \begin{equation}
\left\langle \psi\left|X_{b}\left(t\right)\right|\psi\right\rangle =\left\langle \psi\left(t\right)\left|X_{b}\right|\psi\left(t\right)\right\rangle =\left\langle \psi\left(t\right)\left|\Lambda X\Lambda\right|\psi\left(t\right)\right\rangle =\left\langle \psi_{b}\left(t\right)\left|X\right|\psi_{b}\left(t\right)\right\rangle \,,\end{equation}
\begin{eqnarray}
& \left\langle \psi\left|X_{tr}\left(t\right)\right|\psi\right\rangle   = \left\langle \psi\left(t\right)\left|X_{tr}\right|\psi\left(t\right)\right\rangle =\left\langle \psi\left(t\right)\left|\Lambda X\left(I-\Lambda\right)+\left(I-\Lambda\right)X\Lambda\right|\psi\left(t\right)\right\rangle &  \nonumber\\
 & =  \left\langle \psi_{b}\left(t\right)\left|X\right|\psi_{f}\left(t\right)\right\rangle +\left\langle \psi_{f}\left(t\right)\left|X\right|\psi_{b}\left(t\right)\right\rangle \,, &\end{eqnarray}
 \begin{equation}
\left\langle \psi\left|X_{f}\left(t\right)\right|\psi\right\rangle =\left\langle \psi\left(t\right)\left|X_{f}\right|\psi\left(t\right)\right\rangle =\left\langle \psi\left(t\right)\left|\left(I-\Lambda\right)X\left(I-\Lambda\right)\right|\psi\left(t\right)\right\rangle =\left\langle \psi_{f}\left(t\right)\left|X\right|\psi_{f}\left(t\right)\right\rangle \,.\end{equation}
 Using the limits of $\vert\psi_{b}(t)\rangle $
and $\vert\psi_{f}(t)\rangle $, Eqs. (\ref{eq:backward_component_limits}) and (\ref{eq:forward_component_limits}),
we obtain that the components of  $X(t)$ satisfy
the following limits \begin{eqnarray}
\lim_{t\to-\infty}(X(t)-X_{b}(t))=0\,, &  & \quad\lim_{t\to\infty}X_{b}(t)=0\,,\\
\lim_{t\to-\infty}X_{tr}(t)=0\,, &  & \quad\lim_{t\to\infty}X_{tr}(t)=0\,,\\
\lim_{t\to-\infty}X_{f}(t)=0\,, &  & \quad\lim_{t\to\infty}(X(t)-X_{f}(t))=0\,.\end{eqnarray}

Eq. (\ref{eq:trans_rep_heisenberg_t}) provides the transition decomposition
in the Heisenberg picture. The observable $X(t)$ decomposes into
a sum of three components, $X_{b}(t)$, $X_{tr}(t)$, and $X_{f}(t)$,
such that $X_{b}(t)$ vanishes in the future time asymptote and is
asymptotic to $X(t)$ in the past time asymptote, $X_{f}(t)$ vanishes
in the past time asymptote and is asymptotic to $X(t)$ in the future
time asymptote, and $X_{tr}(t)$ is transient and vanishes both in
the past and in the future time asymptotes. The evolution of $X(t)$
is represented as a transition from $X_{b}(t)$ in the backward asymptote
to $X_{f}(t)$ in the forward asymptote. Accordingly, $X_{b}(t)$ will
be termed the \emph{backward asymptotic transition observable}, $X_{f}(t)$
will be termed the \emph{forward asymptotic transition observable},
and $X_{tr}(t)$ will be termed the \emph{transient observable}.

\section{Application to scattering problems}

In this section we apply the transition decomposition to a quantum
mechanical scattering problem. In particular, we find that this decomposition
is especially useful for the description of scattering resonances.
Specifically, we shall apply the transition decomposition to the evolution
of approximate resonance states defined in the context of the formalism
of \emph{the semigroup decomposition of resonance evolution} \cite{Yossi 05, approx}. This formalism, developed in recent years out of efforts
to adapt the Lax-Phillips scattering theory \cite{Lax} to
the description of the evolution of scattering resonances in quantum
mechanics \cite{quantum Lax}, is a mathematical framework serving as a basis for a time-dependent theory of resonances in quantum mechanical scattering problems \cite{Yossi 03}.
We give here a short presentation of the simplest form of this framework
that is sufficient for our purposes.\\

Recall that Theorem 1 defines a large class of Lypaunov operators
in the sense that for any complete set of generalized eigenstates
of the Hamiltonian there corresponds a different Lyapunov operator.
It follows that each choice of Lyapunov operator has associated with
it its own transition decomposition. We would like to apply 
the transition decomposition to scattering problems. For such problems
there are at least two distinguished energy representations for the
Hamiltonian, i.e. the incoming and outgoing energy representations,
and we must choose which of the two to employ. The incoming and outgoing
energy representations are defined in terms of the incoming and outgoing
solutions of the Lipmann-Schwinger equation. Let $\vert E^{+},\,\lambda\rangle$
and $\vert E^{-},\,\lambda\rangle$ be the incoming and
outgoing solutions of the Lipmann-Schwinger equation satisfying $H\vert E^{\pm},\,\lambda\rangle=E\vert E^{\pm},\,\lambda\rangle$,
where $\lambda$ stands for the degeneracy indices of the energy spectrum.
(The degeneracy is assumed to be uniform over the energy
spectrum.) The \emph{outgoing Lyapunov operator $M^{+}$} is defined
as\begin{equation}
M^{+}=-\frac{1}{2\pi i}\sum_{\lambda}\intop_{0}^{\infty}dE\intop_{0}^{\infty}dE'\vert E^{-},\,\lambda\rangle\frac{1}{E-E'+i0^{+}}\langle E'^{-},\,
\lambda\vert\,
,\end{equation} and gives rise to an \emph{outgoing transition decomposition in the
Schr\"{o}dinger or Heisenberg pictures.} Defining 
$\Lambda^{+}:=\left.M^{+}\right.^{1/2}$, 
 the outgoing transition decomposition 
in the Schr\"{o}dinger picture is given by \begin{equation}
\vert \psi\left(t\right)\rangle =\vert \psi^{+}_{b}\left(t\right)\rangle +\vert \psi^{+}_{f}\left(t\right) \rangle \,,\end{equation}
while for an observable $X$ defined on $\mathcal{H}$, in the outgoing
transition decomposition in the Heisenberg picture, we have \begin{equation}
X=X^{+}_{b}+X^{+}_{tr}+X^{+}_{f}\,,\label{eq:outgoing_trans_rep_heisenberg}\end{equation}
with \begin{equation}
X^{+}_{b}  = \Lambda^{+}X\Lambda^{+}\,,\qquad X^{+}_{tr} =  \Lambda^{+}X(I-\Lambda^{+})+(I-\Lambda^{+})X\Lambda^{+}\,,\qquad X^{+}_{f} = (I-\Lambda^{+})X(I-\Lambda^{+})\,.\end{equation}

To proceed we assume the following:

\renewcommand{\labelenumi}{(\roman{enumi})}

\begin{enumerate}
\item Let $\mathcal{H}$ be a Hilbert space corresponding to
a given quantum mechanical scattering problem. A self-adjoint
`free' unperturbed Hamiltonian $H_{0}$
and a self-adjoint perturbed Hamiltonian $H$ are defined on $\mathcal{H}$
and form a complete scattering system, i.e., we assume that the M\o
ller wave operators $\Omega^{\pm}(H_{0},\, H)$ exist and are complete.

\item We assume that the (absolutely) continuous spectrum
of $H$ is uniformly degenerate. To simplify matters we assume that
this degeneracy is one.

\item The $S$-matrix in the the energy representation,
denoted by $\tilde{\mathcal{S}}(E)$, is the boundary value of a function ${\mathcal S}(z)$ analytic in some strip above the positive real axis and having 
an analytic continuation across the cut on the positive real energy axis into some simply-connected region $\Sigma$ below the real axis in which it has a single, simple resonance pole
at the point $z=\mu$ with $\mathrm{Im\mu<0}$.
\end{enumerate}

It is shown in \cite{Yossi 05, approx} that there exists a dense set $\Xi\subset\mathcal{H}$
and a well defined state $\vert\psi_{\mu}^{app}\rangle \in\mathcal{H}$
such that for any states $\vert\varphi\rangle\in\Xi$ and $\vert\psi\rangle\in\mathcal{H}$
the above assumptions lead to a decomposition, induced by
the pole of the $S$-matrix at $z=\mu$, of matrix elements of the evolution
$U(t)$ generated by $H$, having the form \begin{equation}
\langle \varphi \mid U(t)\psi\rangle=B(\varphi,\psi,\mu;t)+\alpha(\varphi,\mu)\langle \psi_{\mu}^{app}\mid\psi\rangle \, e^{-i\mu t}\,,\qquad t\geq0\,.\label{eq:semigroup_decomp}\end{equation}
The second term of the decomposition on the right-hand side of Eq.
(\ref{eq:semigroup_decomp}) exhibits the typical decay behavior of
a resonance. This term is the \emph{semigroup term}, or \emph{resonance
term}. The term $B(\varphi,\psi,\mu;t)$ in Eq. (\ref{eq:semigroup_decomp})
is the so-called \emph{background} term. The state $\vert\psi_{\mu}^{app}\rangle $
appearing in the second term in Eq. (\ref{eq:semigroup_decomp}) is
called the \emph{approximate resonance state}. Note that if the state
$\vert\psi\rangle$ on the left-hand side of Eq. (\ref{eq:semigroup_decomp})
is chosen to be orthogonal to $\vert\psi_{\mu}^{app}\rangle$ then the resonance
term in that equation vanishes. The reference to $\vert\psi_{\mu}^{app}\rangle$
as an approximate resonance state stems from the fact that it can
be shown that there is no choice of $\vert\varphi\rangle$ and $\vert\psi\rangle$ in the
matrix element $\langle\varphi\mid U(t)\psi\rangle$ for which the background term
$B(\varphi,\psi,\mu;t)$ disappears \cite{Yossi 05}. Indeed, the Schr\"{o}dinger evolution of a closed system does not allow for an exponential decay law for the survival probability \cite{Larry}
and deviations, such as the Zeno effect for short times are inevitable \cite{Zeno}.

An explicit expression for the approximate resonance state $\vert\psi_{\mu}^{app}\rangle$
is given by \cite{Yossi 05, approx, Yossi 05 2}  \begin{equation}
\vert\psi_{\mu}^{app}\rangle =\frac{1}{2\pi i}\int_{\mathbb{R}^{+}}dE\,\frac{1}{E-\mu}\,\vert E^{-}\rangle\,.\label{eq:psi_approx}\end{equation}
This expression is obtained under the assumption
that there is only a single resonance pole of the $S$-matrix below
the positive real axis in the region $\Sigma$. In the case that $\Sigma$
contains multiple resonance poles of the $S$-matrix Eq. (\ref{eq:psi_approx})
provides only a \emph{zeroth order approximate resonance state} \cite{approx}.
Throughout the rest of the paper we only consider approximate resonance
states given by Eq. (\ref{eq:psi_approx}). This restriction pertains
also to the example worked out below 
for which the $S$-matrix possesses multiple resonance poles. Hence,
all states calculated there are of zeroth order. The restriction
to zeroth order approximate resonance states is made for the sake of simplicity and clarity
of exposition and it should be emphasized that there is no a priori
difficulty in working with higher order approximate resonance states.

Applying the outgoing transition decomposition in Eq. (\ref{eq:outgoing_trans_rep_heisenberg})
to the expectation value of an observable $X(t)$ in the state
$\vert\tilde{\psi}_{\mu}^{app}\rangle: = \Vert\vert\psi_{\mu}^{app}\rangle\Vert^{-1} \vert\psi_{\mu}^{app}\rangle $,
we get

\begin{equation}
\bigl\langle \tilde{\psi}_{\mu}^{app}\bigl\vert X(t)\bigr\vert\tilde{\psi}_{\mu}^{app}\bigr\rangle = \bigl\langle \tilde{\psi}_{\mu}^{app}\left\vert X^{+}_{b}(t)\right\vert\tilde{\psi}_{\mu}^{app}\bigr\rangle +\bigl\langle \tilde{\psi}_{\mu}^{app}\bigl\vert X^{+}_{tr}(t)\bigr\vert\tilde{\psi}_{\mu}^{app}\bigr\rangle +\bigl\langle \tilde{\psi}_{\mu}^{app}\bigl\vert X^{+}_{f}(t)\bigr\vert\tilde{\psi}_{\mu}^{app}\bigr\rangle \,.\label{eq:psi_approx_trans_rep}\end{equation}
We shall use this decomposition to represent the evolution of a particular
resonance in a simple one-dimensional scattering problem. The model
we consider is the scattering along
the half-line $\mathbb R^+$ off a square barrier potential. Thus,
we consider a free Hamiltonian $H_{0}=-\frac{1}{2m}\partial_{x}^{2}$ acting on
$L^{2}(\mathbb{R}^{+})$ (where $H_{0}$ is taken to be the self-adjoint
extension in $L^{2}(\mathbb{R}^{+})$ of $-\frac{1}{2m}\partial_{x}^{2}$ from
its original domain of definition $\mathcal{D}(\partial_{x}^{2})=\{\phi(x)\mid\phi(x)\in W_{2}^{2}(\mathbb{R}^{+}),\ \phi(0)=0\}$)
and a full Hamiltonian $H=H_{0}+V$ with $V$ a multiplicative
operator $(V\psi)(x)=V(x)\psi(x)$ such that \begin{equation}
V(x)=\begin{cases}
0\,, & 0<x<a\,,\\
V_{0}\,, & a\leq x\leq b\,,\\
0\,, & b<x\,,\end{cases}\end{equation}
 where $0<a<b$ and $V_{0}>0$. In this case there are no bound state
solutions of the eigenvalue problem for $H$ and the (absolutely)
continuous spectrum of $H$ is $\mathbb{R}^{+}$. In order to find
the scattering states and calculate the $S$-matrix 
one solves the eigenvalue problem for the continuous spectrum generalized
eigenfunctions $\psi_{E}(x)$ of $H$ \begin{equation}
\left(-\frac{1}{2m}\partial_{x}^{2}+V(x)\right)\psi_{E}(x)=E\,\psi_{E}(x)\,,\qquad E\in\mathbb{R}^{+}\,.\end{equation}
 Imposing appropriate boundary conditions we find that \begin{equation}
\psi_{E}(x)=\begin{cases}
\alpha_{1}(k)\sin kx\,, & 0<x\leq a\,,\\
\alpha_{2}(k)e^{ik'x}+\beta_{2}(k)e^{-ik'x}\,, & a<x<b\,,\\
\alpha_{3}(k)e^{ikx}+\beta_{3}(k)e^{-ikx}\,, & b\leq x\,,\end{cases}\label{eq:E_general_eigenfunc}\end{equation}
 where $k=\sqrt{2mE}$ and $k'=\sqrt{2m(E-V_{0})}$ for $E\geq V_{0}>0$
or $k'=i\sqrt{2m{V_{0}-E}}$ for $V_{0}>E\geq0$. The coefficients in
Eq. (\ref{eq:E_general_eigenfunc}) are given by \cite{MG} \begin{equation}
\begin{split}\alpha_{2}(k) & =\frac{1}{2}e^{-ik'a}\left(\sin ka+\frac{k}{ik'}\cos ka\right)\alpha_{1}(k)\,,\\
\beta_{2}(k) & =\frac{1}{2}e^{ik'a}\left(\sin ka-\frac{k}{ik'}\cos ka\right)\alpha_{1}(k)\,,\\
\alpha_{3}(k) & =\frac{1}{4}e^{-ikb}\left(\Bigl(1+\frac{k'}{k}\Bigr)e^{ik'(b-a)}\bigl(\sin ka+\frac{k}{ik'}\cos ka\bigr)\right.\\
 & +\left.\Bigl(1-\frac{k'}{k}\Bigr)e^{-ik'(b-a)}\Bigl(\sin ka-\frac{k}{ik'}\cos ka\Bigr)\right)\alpha_{1}(k)\,,\\
\beta_{3}(k) & =\alpha_{3}^{*}\left(k\right)\,,\end{split}
\label{eq:k_coeff}\end{equation}
 where $\alpha_{1}(k)$ is to be determined by normalization conditions. 
 
Given the full set of solutions $\{\psi_{E}(x)\}_{E\in\mathbb{R}^{+}}$
for the continuous energy spectrum one can find the sets $\{\psi_{E}^{\pm}\}_{E\in\mathbb{R}^{+}}$
of solutions of the Lippmann-Schwinger equation corresponding to incoming
and outgoing asymptotic conditions. We have \begin{equation}
 \psi_{E}^{+}(x) := \langle x\vert E^{+}\rangle =\frac{i}{2\alpha_{3}^{*}(k)}\psi_{E}(x)\,,\qquad \psi_{E}^{-}(x)  :=  \langle x\vert E^{-}\rangle = \frac{1}{2i\alpha_{3}(k)}\psi_{E}(x)\,.\label{eq:lippmann_schwinger_sol}\end{equation}
 where $\psi_{E}^{+}(x)$ and $\psi_{E}^{-}(x)$ are, respectively,
the incoming and outgoing Lippmann-Schwinger solutions. The normalization
conditions for the Lippmann-Schwinger states in Eq. (\ref{eq:lippmann_schwinger_sol})
give us $\alpha_{1}(k)=(2\pi k)^{-1/2}$. In the energy representation
the $S$-matrix is given by \begin{equation}
\tilde{\mathcal{S}}(E)=-\frac{\alpha_{3}(k)}{\alpha_{3}^{*}(k)}\,.\end{equation}
 The above expression for the $S$-matrix leads to the calculation
of the scattering resonances of the problem. For a resonance point
$z=\mu_{j}$ in the lower half-plane below the positive real axis
we set $\mu_{j}=E_{\mu_{j}}-i\Gamma_{\mu_{j}}/2$ with $E_{\mu_{j}}>0$
the resonance energy and $\Gamma_{\mu_{j}}>0$ the resonance width. 
\par For barrier parameters $a=2m^{-1}$, $b=3m^{-1}$, and $V_{0}=5 m$ the three lowest
energy resonance poles are given by $\mu_{1}\simeq 0.9106\,m-i\,0.0012\,m$,
$\mu_{2}\simeq3.5119\,m-i\,0.0282\,m$ and $\mu_{3}=7.1168\,m-i\,0.4462\,m$. We shall
focus on the third resonance pole $\mu_{3}$. Utilizing the outgoing
Lippmann-Schwinger eigenfunctions $\psi_{E}^{-}(x)$, given by Eqs.
(\ref{eq:E_general_eigenfunc}-\ref{eq:lippmann_schwinger_sol}),
the spatial wave function of the approximate resonance state $\psi_{\mu_{3}}^{app}(x)$, and its energy density,
can be calculated numerically. The probability density $\vert\psi_{\mu_{3}}^{app}(x)\vert^{2}$
is shown in Fig. 1 (see \cite{approx}), while
the energy density $\vert\psi_{\mu_{3}}^{app}(E)\vert^{2}$ is shown in Fig. 2.

\begin{figure}[t]
\begin{minipage}[b]{0.475\linewidth}
\centering
\includegraphics[scale=0.6]{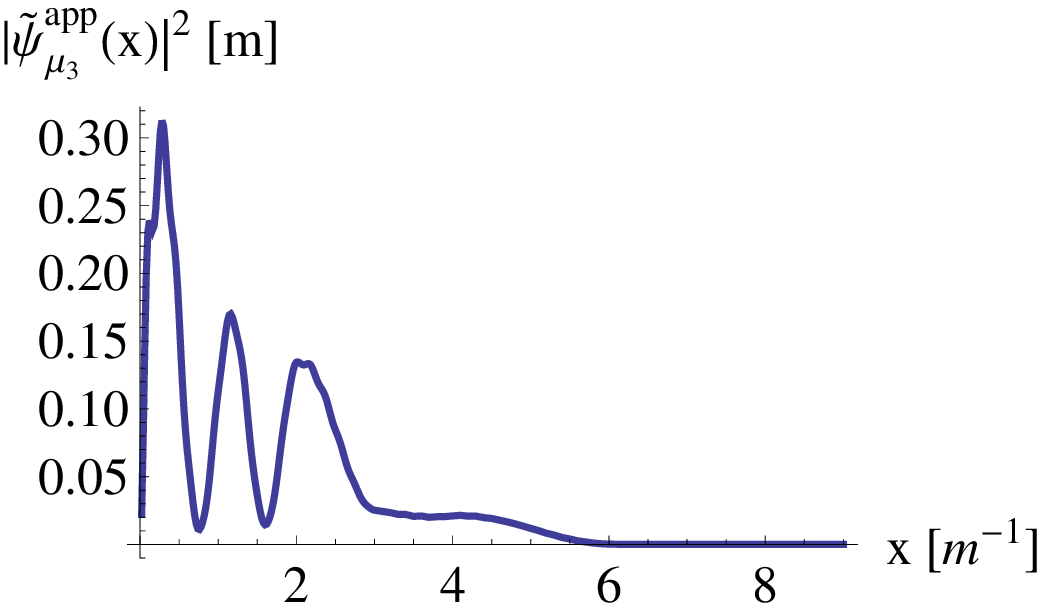}
\caption{Spatial density of the third 
resonance $\mu_3$.}

\end{minipage}
\hspace{0.5cm}
\begin{minipage}[b]{0.475\linewidth}
\centering
\includegraphics[scale=0.6]{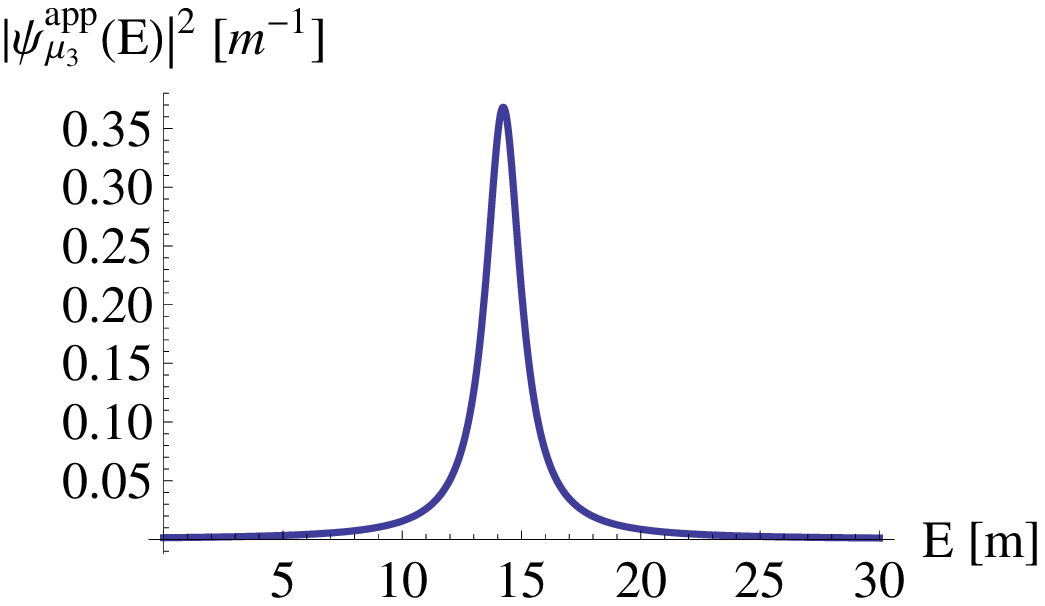}
\caption{Energy density of the third resonance $\mu_3$. }

\end{minipage}
\end{figure}

Now apply the outgoing transition decomposition of Eq. (\ref{eq:psi_approx_trans_rep})
to the evolution of the approximate resonance state $\vert\psi_{\mu_{3}}^{app}(t)\rangle =U(t)\vert\psi_{\mu_{3}}^{app}\rangle $.
According to Eq. (\ref{eq:psi_approx_trans_rep}) the spatial probability distribution
of $\vert\psi_{\mu_{3}}^{app}\left(x,\, t\right)\vert^{2}$ decomposes
into three components\begin{equation}
\label{eq:trans_rep_prob_density}
\bigl\vert\tilde{\psi}_{\mu_{3}}^{app}\left(x,\, t\right)\bigr\vert^{2}=\bigl\vert\tilde{\psi}_{\mu_{3},\,b}^{app,\, +}\left(x,\, t\right)\bigr\vert^{2}+2\mathrm{Re}\bigl(\bigl(\tilde{\psi}_{\mu_{3},\,b}^{app,\, +}\left(x,\, t\right)\bigr)^*\tilde{\psi}_{\mu_{3},\,f}^{app,\, +}\left(x,\, t\right)\bigr)+\bigl\vert\tilde{\psi}_{\mu_{3},\,f}^{app,\, +}\left(x,\, t\right)\bigr\vert^{2}\,.
\end{equation}
\begin{figure}[t]
\center{ \includegraphics[scale=0.6]{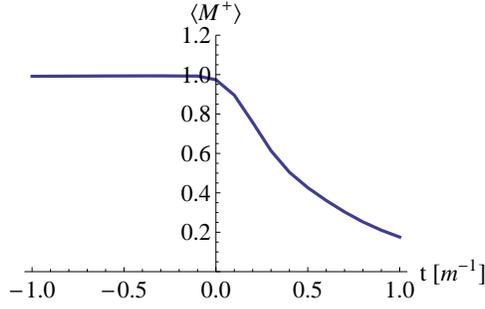}
\caption{Monotonic decrease of the expectation value of $M^+$ in the state $\vert\tilde{\psi}_{\mu_3}^{app}(t)\rangle$. Note the transition to exponential decay for $t>0$.}}
\end{figure}The right-hand side of Eq. (\ref{eq:trans_rep_prob_density}) is the
outgoing transition decomposition of the position probability density
of $\psi_{\mu_{3}}^{app}(t)$. The first term on the right hand side
of Eq. (\ref{eq:trans_rep_prob_density}) is the backward asymptotic
component, the second term is the transient component, and the third
term is the forward asymptotic component.
\begin{figure}
\setlength{\unitlength}{1cm}
\vskip -3cm
\hskip -15.0cm
\begin{minipage}{5cm}
\mbox{\includegraphics[height=30cm, width=22cm]{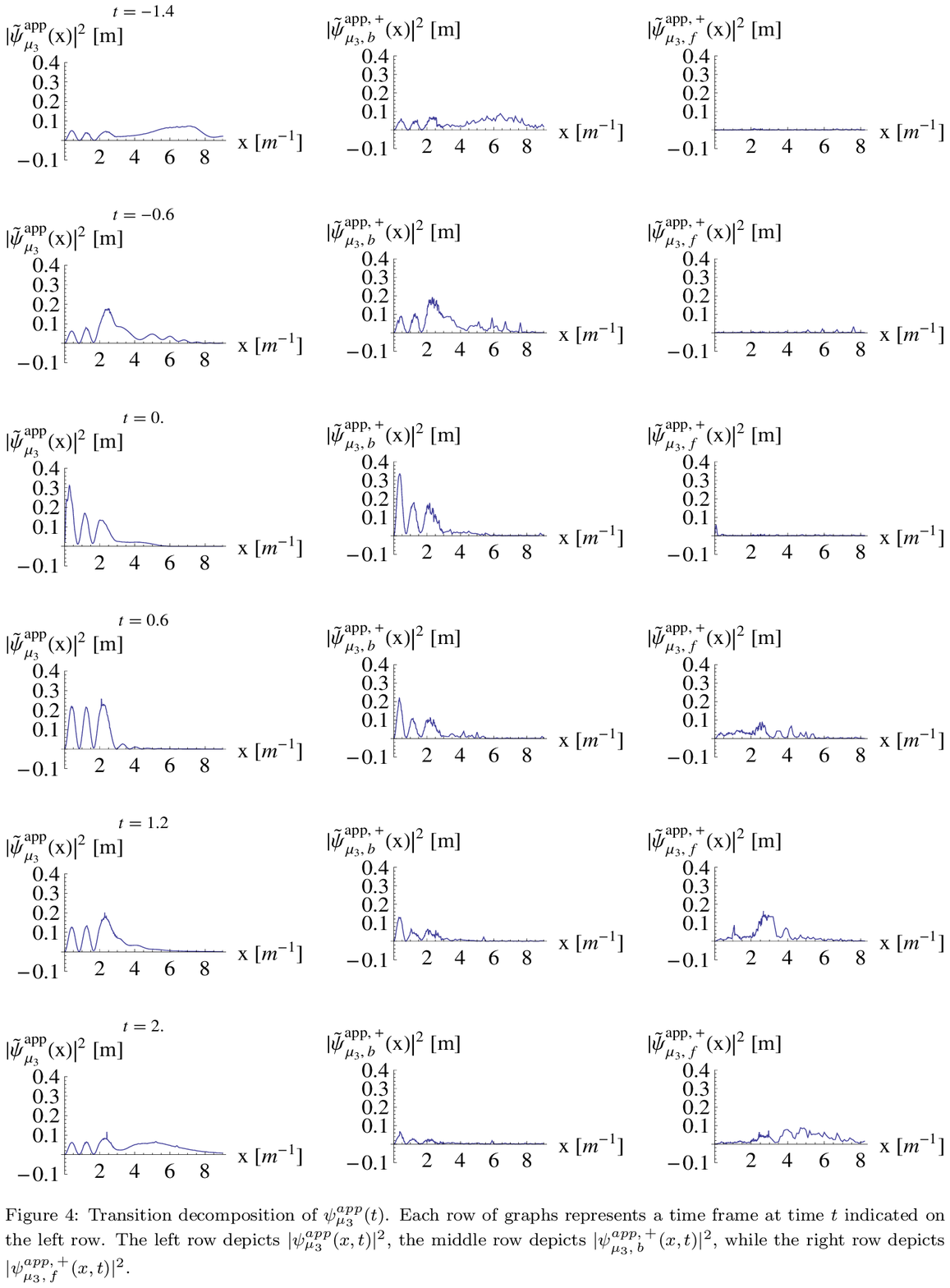}}
\end{minipage}
\end{figure}
Fig. 4 presents the results of the
application of the outgoing transition decomposition to the evolution
of the approximate resonance state $\vert\psi_{\mu_{3}}^{app}\rangle $.
Each row represents a `snapshot' corresponding to a particular time $t$. The left graph in each row shows $\vert\tilde{\psi}_{\mu_{3}}^{app}(x,\, t)\vert^2$ at time $t$. The middle
graph shows the contribution of its backward asymptotic component $\vert\tilde{\psi}_{\mu_{3},\,b}^{app,\,+}(x,\, t)\vert^{2}$
at time $t$, and the right graph in each row shows the sum of transient and forward asymptotic components  $2\mathrm{Re}((\tilde{\psi}_{\mu_{3},\,b}^{app,\, +}(x,\, t))^{*}\tilde{\psi}_{\mu_{3},\,f}^{app,\, +}(x,\, t))+\vert\tilde{\psi}_{\mu_{3},\,f}^{app,\, +}(x,\, t)\vert^{2}$. It is clearly seen from this sequence of snapshots 
that the formation phase of the resonance, starting at the negative
time asymptote $t\to-\infty$ and following through the scattering
process up to $t=0$, is captured by the backward asymptotic component 
(the middle column of graphs),
while through this whole time interval the contribution of the transient and forward asymptotic components is small. In the decay phase of the resonance,
commencing at $t>0$, the middle column of graphs essentially shows the spatial
probability density profile of a resonance state, multiplied by an
exponentially decaying factor $\exp(-\Gamma_{\mu_{3}}t)$, which gives the
decay of the resonance over time. As the resonance state decays the
probability is transferred to the forward asymptotic term, a process
captured in the right most column of graphs, 
and is eventually carried to spatial infinity as the scattering process
evolves further towards the forward time asymptote $t\to\infty$.
Observe also that the graphs on the right column  in Fig. 4
obtain also negative values. This is due to the contribution of the
transient term which is not necessarily positive. 

Further understanding of the behavior of the probability density $\vert\tilde{\psi}_{\mu_{3},\,b}^{app,\,+}(x,t)\vert^{2}$
may be gained by integrating it over $x$. In this way we get \begin{eqnarray}
 &  \, & \intop_{0}^{\infty}dx\,\bigl\vert\tilde{\psi}_{\mu_{3},\,b}^{app,\,+}(x,\, t)\bigr\vert^{2}\;=\;\intop_{0}^{\infty}dx\bigl\langle \tilde{\psi}_{\mu_{3},\,b}^{app,\,+}\left(t\right)\mid x\bigr\rangle \bigl\langle x\mid\tilde{\psi}_{\mu_{3},\,b}^{app,\,+}\left(t\right)\bigr\rangle  \nonumber \\
 &  = & \intop_{0}^{\infty}dx\bigl\langle \Lambda^{+}\tilde{\psi}_{\mu_{3}}^{app}\left(t\right)\mid x\bigr\rangle \bigl\langle x\mid\Lambda^{+}\tilde{\psi}_{\mu_{3}}^{app}\left(t\right)\bigr\rangle \;= \;\bigl\langle \tilde{\psi}_{\mu_{3}}^{app}\left(t\right)\left|M^{+}\right|\tilde{\psi}_{\mu_{3}}^{app}\left(t\right)\bigr\rangle\,,  \end{eqnarray}
which is just the expectation value of the Lyapunov operator $M^{+}$ in
the state $\left|\psi_{\mu_{3}}^{app}(t)\right\rangle $. A plot
of this behavior is presented in Fig. 3. The decay of the expectation
value of $M^{+}$ matches the resonance decay factor $\exp(-\Gamma_{\mu_{3}}t)$
mentioned above.

\newpage

\begin{acknowledgments}

Y. Strauss, and J. Silman and S. Machnes, acknowledge support from the ISF (Grants. 1169/06 and 784/06, respectively). J. Silman also acknowledges the support of the Inter-University Attraction
Poles Programme (Belgian Science Policy) under Project IAP-P6/10 (Photonics@be) and of the FNRS.

\end{acknowledgments}

\end{document}